\documentclass[onecolumn,aps,prb, superscriptaddress]{revtex4-2}
\usepackage[normalem]{ulem}
\usepackage{epsfig}
\usepackage{color}
\usepackage{amsmath}
\usepackage{enumitem}
\usepackage{hyperref}
\hypersetup{
	colorlinks = true,
	linkcolor = blue,
	citecolor = blue,
	urlcolor = blue
}
\usepackage[normalem]{ulem}
\newcommand{\be}{\begin{equation}}
\newcommand{\ee}{\end{equation}}

\newcommand{\bea}{\begin{eqnarray}}
\newcommand{\eea}{\end{eqnarray}}

\newcommand{\sign}{{\rm sign\,}}
\newcommand{\tr}{{\rm \, tr\,}}

\renewcommand{\vec}[1]{{\bf #1}}
\renewcommand{\phi}{\varphi}
\renewcommand{\epsilon}{\varepsilon}
\def\nn{\nonumber\\}

\renewcommand{\cite}[1]{[\onlinecite{#1}]}

\numberwithin{equation}{section}
\begin{document}

\title{Gap generation and flat band catalysis in dice model with local interaction}

\author{E. V. Gorbar}
\affiliation{Department of Physics, Taras Shevchenko National University of Kyiv, Kyiv, 03022, Ukraine}
\affiliation{Bogolyubov Institute for Theoretical Physics, Kyiv, 03143, Ukraine}

\author{V. P. Gusynin}
\affiliation{Bogolyubov Institute for Theoretical Physics, Kyiv, 03143, Ukraine}

\author{D. O. Oriekhov}
\affiliation{Instituut-Lorentz, Universiteit Leiden, P.O. Box 9506, 2300 RA Leiden, The Netherlands}

\begin{abstract}

The gap generation in the dice model with local four-fermion interaction is studied. Due to the presence of two valleys with degenerate electron states, there are
two main types of gaps. The intra- and intervalley gap describes the electron and hole pairing in the same and different valleys, respectively. We found that
while the generation of the intravalley gap takes place only in the supercritical regime, the intervalley gap is generated for an arbitrary small coupling.
The physical reason for the absence of the critical coupling is the catalysis of the intervalley gap generation by the flat band in the electron spectrum of the dice model. The completely quenched
kinetic energy in the flat band when integrated over momentum in the gap equation leads to extremely large intervalley gap proportional to the area
of the Brillouin zone.

\end{abstract}
\maketitle

\section{Introduction}

The experimental discovery of graphene [\onlinecite{Geim_graphene}] draw attention of condensed matter physicists to the systems with relativisticlike
quasiparticle spectrum. It was shown [\onlinecite{Bradlyn}] that in crystals with special space groups more complicated electron spectra could be realized with no analogues
in high-energy physics where the Poincare symmetry provides strong restrictions. One remarkable example is a possibility to
possess strictly flat bands [\onlinecite{Heikkila1,Heikkila2}], whose high degeneracy was shown to be stabilized by
permutation symmetries \cite{Lima2020PRB} (for a recent review of artificial flat band systems, see Ref.[\onlinecite{Leykam}] and Ref.\cite{Lima2021NanoSc} where many systems
with pseudospin-1 fermions have been discussed). The
dice model is the paradigmatic example of such a system with a flat band which hosts pseudospin-1 fermions [\onlinecite{Sutherland}].

The dice model is a tight-binding model of two-dimensional fermions living on the ${\cal T}_3$ (or dice) lattice where atoms are situated both at the vertices
of hexagonal lattice and the hexagons centers [\onlinecite{Sutherland,Vidal}]. Since the dice model has three sites per unit cell, the electron states in
this model are described by three-component fermions. It is natural then that the energy spectrum of the model is comprised of three bands. The two of them form
a Dirac cone and the third band is completely flat and has zero energy [\onlinecite{Raoux}]. All three bands meet at the $K$ and $K^{\prime}$ points, which are situated at the corners of the Brillouin zone. The ${\cal T}_3$ lattice has been experimentally realized in Josephson arrays [\onlinecite{Serret,Abilio1999}], metallic wire networks [\onlinecite{Naud}] and its optical realization by laser beams was proposed in Ref.[\onlinecite{Rizzi}].

Perfectly flat bands are expected to be unstable with respect to generic perturbations such as the presence
of boundaries, magnetic field, Coulomb impurities, and disorder. In a recent paper [\onlinecite{Oriekhov}], we showed that, remarkably, the energy dispersion of
the completely flat energy band of the dice model is not affected by the presence of boundaries except the trivial reduction of the number of degenerated electron states due to the finite spatial size of the system. It was shown also that the flat band of the dice model remains unaltered in the presence of circularly
polarized radiation \cite{Dey,Iurov_circ} and magnetic field [\onlinecite{Bercioux}]. The electron states of gapped pseudospin-1 fermions in the dice model for
impurities with short- and long-range potential were studied by us in Ref.[\onlinecite{Gorbar2019PRB}] leading to qualitatively different results. Indeed, it was found that while the flat band survives in the 
presence of a potential well, it is absent in the case of the Coulomb potential.

It is well known that a soft kinetic spectrum favors the gap generation. For example, the low energy electron spectrum $\varepsilon(\mathbf{p}) \sim |\mathbf{p}|^n$ in ABC-stacked multilayer graphene becomes more flat with $n$. The interaction parameter $r_s$, defined as the ratio of inter-electron Coulomb interaction energy to the Fermi energy, scales like $r_s\sim n^{(1-n)/2}_{el}$ \cite{Sarma}, where $n_{el}$ is the electron charge density. Obviously, the electron-electron interactions become more important at low electron density as the number of layers $n$ increases in ABC-stacked multilayer graphene. This
suggests that the gap generation in chiral
multilayer graphene should be enhanced \cite{Polini,Sun1,Sun2} as the number of layers $n$ becomes larger. This conclusion agrees with the experimental findings. Meanwhile no gap is observed in monolayer graphene at the neutrality point in the absence of external electromagnetic fields, gap $2$ meV is reported in bilayer graphene \cite{Martin,Weitz,Freitag,Velasco}. The recent experiments \cite{Lee,LeRoy} demonstrate the presence of gaps of almost room temperature magnitude
$\sim 25$ meV in high mobility ABC-stacked trilayer graphene. A large interaction-induced transport gap up to 80 meV was quite recently observed experimentally in suspended rhombohedral-stacked tetralayer graphene \cite{Myhro}.

Obviously, the flat band represents the most extreme case of a soft kinetic spectrum where the kinetic energy is completely quenched. The above mentioned experimental results suggest that the generated gap should have the largest magnitude in the flat band system. This motivates us to study the gap generation in the dice model. A recent theoretical study of the band structure of magic angle twisted bilayer graphene \cite{Cea2020PRB} also shows the crucial role of the flat band and the possibility of large gap generation. This provides an additional motivation for the present study. We would like to add also that since the pseudospin-1 fermions with flat band were already realized in kagome metals such as FeSn \cite{Kang2019} and in electronic Lieb lattice \cite{Slot2017}, our results for the flat band catalysis of gap generation can be tested experimentally.

To get an insight into the gap generation in the dice lattice we considered in this paper a model with local interaction. We studied both intravalley and intervalley types of gap and analyzed their free
energies.

The paper is organized as follows. The dice model and its general properties are considered in Sec.\ref{sec:model}. In Sec.\ref{sec:intravalley}, we study the intravalley gap generation. The intervalley gap generation is
investigated in Sec.\ref{section-intervalley}. In Sec.\ref{intervalley-numerical}, we calculate the free energy for intra- and intervalley gap states and discuss the phase diagram of the model. Technical details of calculations are presented in Appendices \ref{GF-intra}, \ref{GF-inter}, \ref{appendix:free-energy}.

\section{Model}
\label{sec:model}

\begin{figure}	\includegraphics[scale=0.35]{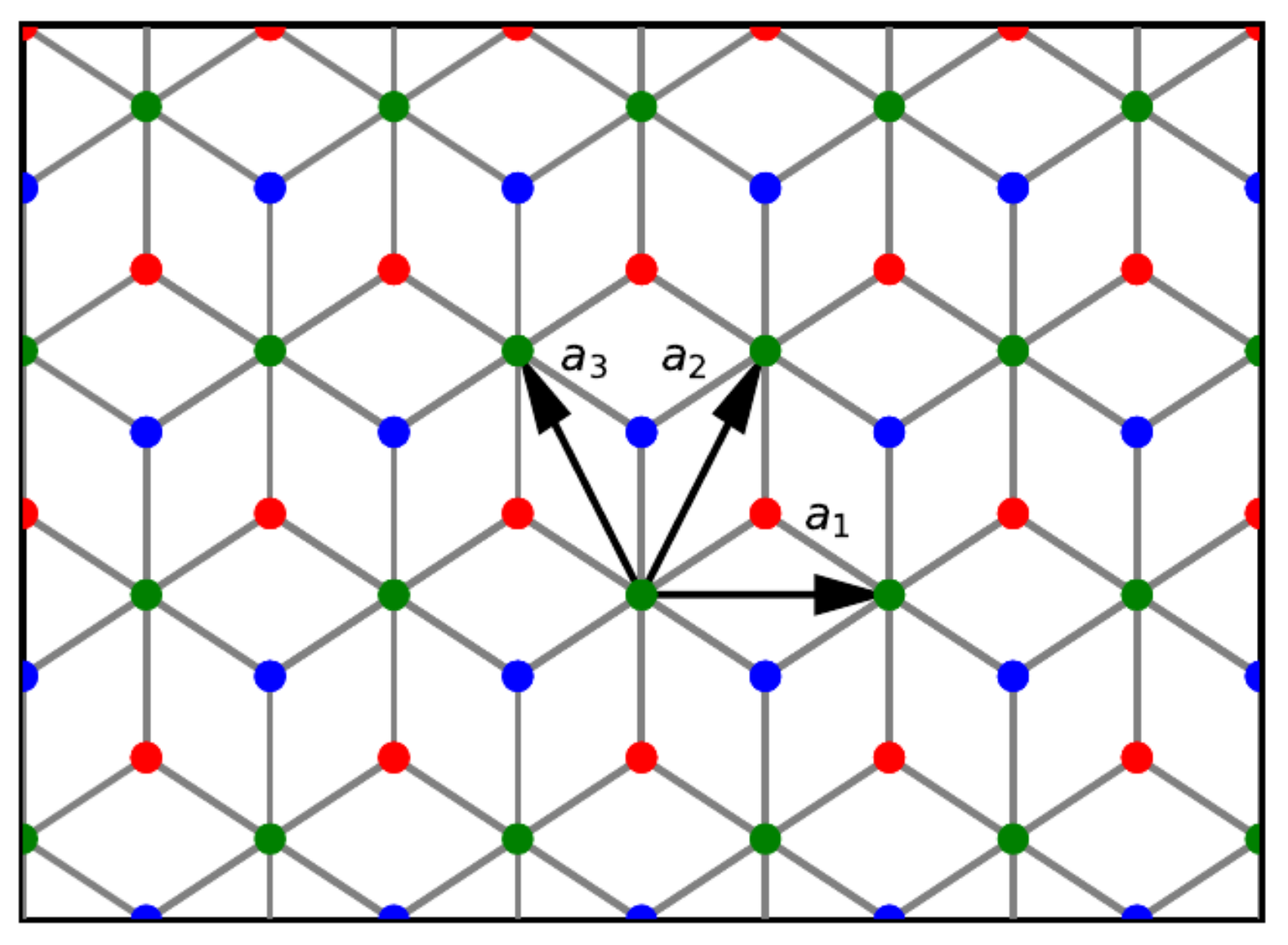}
\caption{A schematic plot of the lattice of the dice model. The red points display the $A$ sublattice atoms, the blue points describe the $B$ sublattice,
and the green points define the $C$ sublattice. The vectors $\vec{a}_1=(1,\,0)a$ and $\vec{a}_2=(1/2,\,\sqrt{3}/2)a$ are the basis vectors of triangular sublattices.}
\label{fig1}
\end{figure}

The lattice of the $\mathcal{T}_{3}$ (dice) lattice model is schematically shown in Fig.\ref{fig1}. The tight-binding equations are [\onlinecite{Bercioux}] (with equal hoppings $t$ between
atoms $C$ and $A,B$)
\begin{align}
&\epsilon\Psi_C(\vec{r})=-t\sum\limits_{j}\Psi_{A}(\vec{r}+\vec{\delta}_{j}^{A})-t\sum\limits_{j}\Psi_{B}(\vec{r}-\vec{\delta}_{j}^{A}),
\nonumber\\
&\epsilon\Psi_{A}(\vec{r})=-t\sum\limits_{j}\Psi_{C}(\vec{r}-\vec{\delta}_{j}^{A}),\nonumber\\
&\epsilon\Psi_{B}(\vec{r})=-t\sum\limits_{j}\Psi_{C}(\vec{r}+\vec{\delta}_{j}^{A}),
\end{align}
where the vectors $\delta_{j}^{A}$ connect nearest neighbor atoms. The corresponding lattice Hamiltonian is expressed through the function $f_{\mathbf{k}}=-\sqrt{2}t(1+e^{-i\mathbf{k}\mathbf{a}_2}+e^{-i\mathbf{k}\mathbf{a}_3})$ and it is not difficult to
find its energy spectrum \cite{Raoux}
\begin{equation}
\epsilon=0,\quad\quad \epsilon=\pm |f_{\mathbf{k}}|=\pm\sqrt{2}t\bigg[3+2(\cos(\vec{a}_1\vec{k})+\cos(\vec{a}_2\vec{k})+\cos(\vec{a}_3\vec{k}))\bigg]^{1/2},
\label{dice-model-energy-spectrum}
\end{equation}
where $\vec{a}_1=(1,\,0)a$ and $\vec{a}_2=(1/2,\,\sqrt{3}/2)a$ are the basis vectors of the triangle sublattices and $\vec{a}_3=\vec{a}_2 - \vec{a}_1$ and $a$ is the lattice constant. The presence of a 
completely flat band with zero energy is perhaps one of the most remarkable properties of the dice model.

There are two values of momentum where $f_{\vec{k}}=0$ and all three bands meet. They are situated at the corners of the hexagonal
Brillouin zone
\begin{align}
	K=\frac{2\pi}{a}\left(\frac{1}{3},\,\frac{1}{\sqrt{3}}\right),\quad K'=\frac{2\pi}{a}\left(-\frac{1}{3},\,\frac{1}{\sqrt{3}}\right).
\end{align}
For momenta near the $K$ and $K^{\prime}$ points, the function $f_{\vec{k}}$ is linear in $\vec{q}=\vec{k}-\xi\vec{K}$, i.e.,
$f_{\vec{k}}=v_F(\xi q_x-iq_y)$, $v_F=\sqrt{3}ta/2$ is the Fermi velocity, and $\xi=\pm$ is the valley index. In addition, we set $\hbar=1$ for convenience.
The low-energy Hamiltonian for electron states of the dice model in both valleys has the form
\begin{align}
H_0(\vec{k},\xi)=
\left(\begin{array}{ccc}
0 & \xi k_x-ik_y & 0\\
\xi k_x+ik_y & 0 & \xi k_x-ik_y\\
0 & \xi k_x+ik_y & 0
\end{array}\right).
\label{Hamiltonian-free}
\end{align}
Here we absorbed dimensional constant $v_F/\sqrt{2}$ into the definition of momenta $\vec{k}=(v_F /\sqrt{2})\,\vec{q}$ (this $\vec{k}$ should not be confused with the initial $\vec{k}$ in the Brillouin zone in
Eq.(\ref{dice-model-energy-spectrum})).
The Hamiltonian acts on three-component wave functions $\psi^T=(\psi_{A},\psi_{C},\psi_{B})$. The
electron states at the $K^{\prime}$ point are described like in graphene by the interchange of the A and B spinor components.
The two valley Hamiltonian, $H_0(\vec{k},+1)\oplus H_0(\vec{k},-1)$,
is time-reversal invariant because of the relation $H^*_0(\vec{k},\xi)=H_0(-\vec{k},-\xi)$, which can be directly checked for Eq.\eqref{Hamiltonian-free}. The time-reversal operator $\mathcal{T}$ for the dice model is defined in the same way as in graphene: it interchanges valleys, changes the sign of momentum, and contains complex conjugation operator \cite{Katsnelson}. The spectrum of the Hamiltonian
consists of three energy bands $\pm \sqrt{2}|\mathbf{k}|,\,0$. Clearly, two bands form a Dirac cone and one band is completely flat.

Although electrons interact through the Coulomb interaction $V(\vec{x}-\vec{y})=e^2/|\mathbf{x}-\mathbf{y}|$, to get an insight into
the gap generation for quasiparticles in the dice model we will study the gap generation for a local Coulomb interaction $V_{local}(\vec{x}-\vec{y})=U\delta^2(\vec{x}-\vec{y})$. This significantly simplifies
the analysis because the gap equations are algebraic in the Hartree--Fock approximation rather than the integral ones as for the genuine Coulomb interaction.
The interaction $V_{local}$ is attractive between electrons and holes. There are two main possibilities of order parameters of the exciton type, namely, the intravalley and intervalley pairing which will be
investigated in the two subsequent sections.

We will study the gap generation by using the Baym--Kadanoff formalism \cite{BK1,BK2,BK3}. The corresponding
effective action for the quasiparticle propagator $G$ in the Hartree--Fock (mean field) approximation in the model with the local four-fermion interaction has
the form (for a similar consideration in the case of graphene, see, e.g., \cite{Gorbar2008PRB})
\begin{equation}
	\Gamma(G)=-\mathrm{i} \operatorname{Tr}\left[\operatorname{Ln} G^{-1}S+\left(S^{-1} G-1\right)\right] + \frac{U}{2}\int d^3x\,\left\{\,\mbox{tr}[G(x,x)G(x,x)] - [\mbox{tr}\,G(x,x)]^2\,\right\},
\label{BK-action}
\end{equation}
where $Tr$ and $Ln$ are taken in the functional sense, $S$ is the free propagator related to Hamiltonian (\ref{Hamiltonian-free}), and trace is taken over the valley and spinor components. Let us begin our analysis with the case of the intravalley gap generation.

\section{Intravalley gap}
\label{sec:intravalley}

First of all, let us consider possible intravalley gap terms in the dice model, whose dynamical generation will be analyzed below.
Obviously, the most general momentum-independent intravalley gap term is given by
\begin{align}
H_{\mbox{\tiny gap}}=\left(\begin{array}{ccc}
m_1 & c & a\\
c^* & m_2 & b\\
a^* & b^* & m_3
\end{array}\right).
\label{mass-terms}
\end{align}
It is easy to check that parameters $a$, $b$, and $c$ lead to an energy dispersion relation which is anisotropic in momentum space. Since it is natural to
expect that the solution with the lowest energy should be isotropic in a rotation-invariant system, we will set $a=b=c=0$ in what follows. Then $m_1$, $m_2$, and $m_3$ are possible mass
terms and. The electron states at the two different valleys are independent, therefore, $m_i$ could be valley dependent (note that valley-polarized states are well-known in graphene \cite{Garcia-Pomar,Abergel,Goerbig}). Next we find the following characteristic equation which determines the energy spectrum of the Hamiltonian $H_0(\mathbf{k},\xi)+H_{\mbox{\tiny gap}}$:
\begin{equation}
(m_1-\epsilon)\left(\,(m_2-\epsilon)(m_3-\epsilon) - k^2\,\right) + (\epsilon-m_3)k^2=0,\quad k=|\mathbf{k}| .
\label{energy-dispersion-isotropic}
\end{equation}
Clearly, there are three solutions of the above equation. Two of them tend to $\epsilon(\mathbf{k}) \to \pm \sqrt{2}\,k$ at $k \to \infty$ and describe
the upper and lower energy branches of the non-perturbated Hamiltonian (\ref{Hamiltonian-free}).
Obviously, if $m_1=-m_3$, then the middle branch tends to the flat energy band $\epsilon=0$ of the free Hamiltonian (\ref{Hamiltonian-free}) at large $|\mathbf{k}|$. Therefore, we will assume in what follows that
$m_1=-m_3=m$. In this case, Eq.(\ref{energy-dispersion-isotropic}) takes the form
\begin{equation}
(\epsilon-m_2)(m^2-\epsilon^2) + 2\epsilon k^2=0.
\label{energy-dispersion-isotropic-1}
\end{equation}

\begin{figure}
	\centering
\includegraphics[scale=0.65]{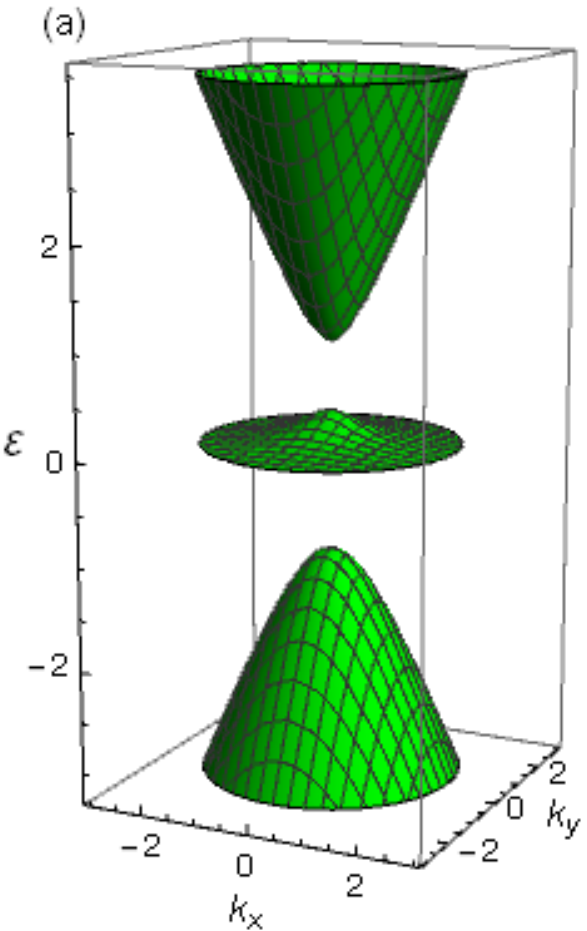}\qquad
\includegraphics[scale=0.65]{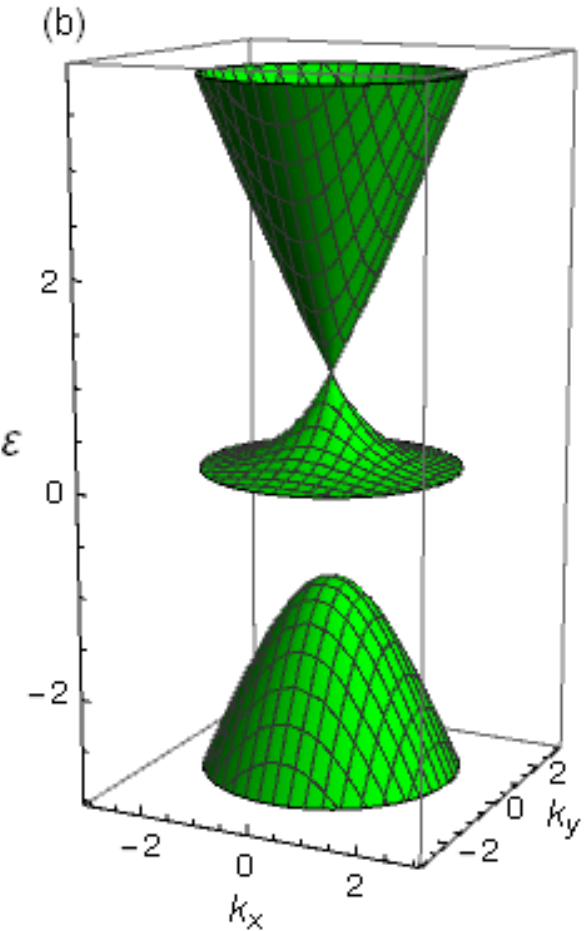}\qquad
\includegraphics[scale=0.65]{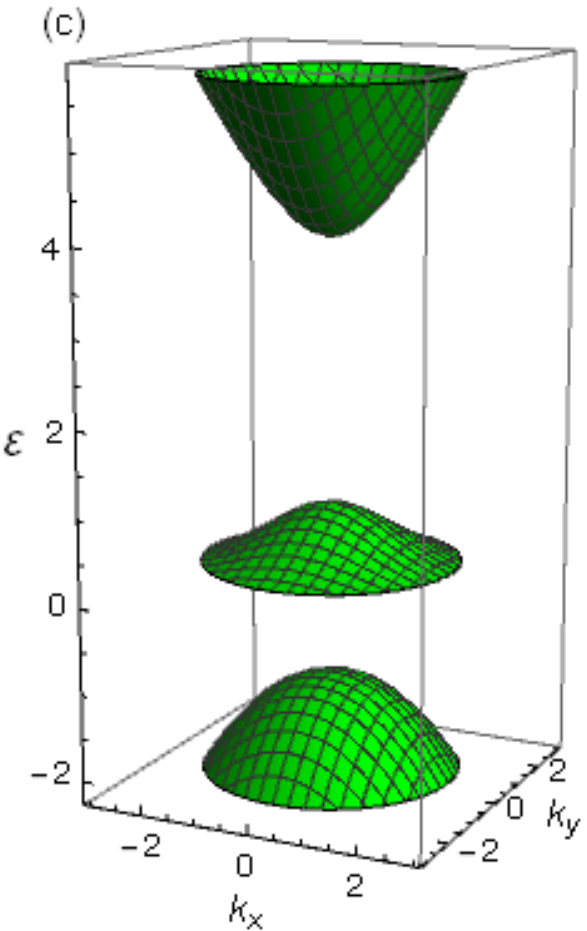}
\caption{Energy spectrum defined by Eq.\eqref{energy-dispersion-isotropic-1} for three values of $m_2$ (a): $m_2=0.35m$, (b): $m_2=m$, (c): $m_2=4m$. At the middle panel the crossing point of two bands is shown. Here
energy $\epsilon$ and momenta $\mathbf{k}$ are measured in units of $m$.}
\label{eq:figure-spectrum}
\end{figure}

The examples of spectrum defined by this equation are shown in Fig.\ref{eq:figure-spectrum}. It is easy to check that $\epsilon=0$ is the exact solution of Eq.(\ref{energy-dispersion-isotropic-1}) for all $\mathbf{k}$ if $m_2=0$. The flat band solution $\epsilon=0$ is realized also if $m=0$. In what follows, we will study only solutions with $ m \ne 0$ and $m_2 \ne 0$ when the flat band is absent. Equation (\ref{energy-dispersion-isotropic-1}) implies that the particle-hole symmetry could be preserved even in the case $m_2 \ne 0$ if we consider the mass term $-m_2$ at the valley $\xi=-$. Since the choice of the sign
of $m$ is irrelevant for the energy dispersion, without loss of generality we can assume that $m$ takes the same value in both valleys. Thus, we have the following intravalley gapped Hamiltonian at
valley $\xi$:
\begin{align}
H_{\xi}=\left(\begin{array}{ccc}
0 & \xi k_x-ik_y & 0\\
\xi k_x+ik_y & 0 & \xi k_x-ik_y\\
0 & \xi k_x+ik_y & 0
\end{array}\right)
+\left(\begin{array}{ccc}
m & 0 & 0\\
0 & \xi m_2 & 0\\
0 & 0 & -m
\end{array}\right).
\label{Hamiltonian-quadratic}
\end{align}
It is worth noting that this Hamiltonian for $m_2=0$ possesses the intravalley particle-hole symmetry ${\cal C}=AK$
\begin{equation}
{\cal C}H_{\xi}(\mathbf{k})+H_{\xi}(\mathbf{k}){\cal C}=0, \quad\quad A=\left(\begin{array}{ccc}
0 & 0 & 1\\
0 & -1 & 0\\
1 & 0 & 0
\end{array}\right),
\label{charge-conjugation}
\end{equation}
where $K$ is the complex conjugation. The relation above can be checked straightforwardly. The existence of this particle-hole symmetry explains why the energy spectrum is particle-hole symmetric in a given valley for $m_2=0$.
The second term in Hamiltonian (\ref{Hamiltonian-quadratic}) defines an ansatz for the full inverse propagator in the theory with the Hamiltonian $H_0+V_{local}$, where gap parameters $m$ and $m_2$ are determined by solving the Schwinger--Dyson equation.

\subsection{Gap equations}

Varying the Baym--Kadanoff action (\ref{BK-action}) with respect to $G$, we obtain the following Schwinger--Dyson equation in the Hartree-Fock (mean field) approximation:
\begin{equation}
G^{-1}_{\xi}(\Omega,\mathbf{p})=S^{-1}_{\xi}(\Omega,\mathbf{p}) - i\frac{2U}{ v_F^2}\int \frac{d\omega d^2 k}{(2\pi)^3}\,G_{\xi}(\omega,\mathbf{k}),
\label{SD-equation-intra}
\end{equation}
where we retained only the exchange contribution because the Hartree contribution vanishes at the neutrality point of the considered particle-hole symmetric
state. Note that $H_{\xi}$ does not mix states from the two valleys, therefore, the Schwinger--Dyson equation (\ref{SD-equation-intra}) for the intravalley gaps
is diagonal in the valley indices. The additional factor $2/v_F^2$ appears due to the definition of $k$ below Eq.(\ref{Hamiltonian-free}).

As was discussed above, we study the gap generation in a neutral particle-hole symmetric system with $m_2$ and $-m_2$ mass terms in the valleys $+$ and $-$, respectively. Therefore, there is no need to introduce the chemical potential. However, the valley dependent chemical potential $\xi \mu_v$ with opposite signs
in the two valleys could be dynamically generated. Hence it should be added to the Hamiltonian $H_{\xi}$. Such chemical potential defines filling at particular valley $\xi$. 
The corresponding gap equations for $m$, $m_2$, and $\mu_v$ are derived in Appendix \ref{GF-intra}. It is useful to perform the Wick rotation $\omega\to i\omega$ in the gap equations (\ref{equation-m-v-2})-(\ref{equation-m2-2}) and integrate over $\omega$ and polar angle $\phi$. Then we obtain the following system of equations for $\mu_v,m$, and $m_2$:
\begin{align}\label{eq:mu-v-euclid}
\mu_v&=\frac{U}{ v_F^2}\int\limits_{0}^{\Lambda} \frac{ k dk}{2\pi}\,\left[\frac{k^2+r_0
	\left(m_2-r_0\right)
	}{\left(r_1-r_0\right) \left(r_0-r_2\right)
	}\text{sign}[\mu_v-r_0]+c.p.\right],\\
\label{eq:m-euclid}
m&=m\frac{U}{ v_F^2}\int\limits_{0}^{\Lambda} \frac{ k dk}{2\pi}\left[\frac{(m_2-r_0)}{(r_0-r_1)(r_0-r_2)}\text{sign}[\mu_v-r_0]+c.p.\right],\\
\label{eq:m2-euclid}
m_2&=-\frac{U}{ v_F^2}\int\limits_{0}^{\Lambda} \frac{ k dk}{2\pi}\,\left[\frac{k^2-m^2+m_2 r_0
	}{\left(r_0-r_1\right)
	\left(r_0-r_2\right) }\text{sign}[\mu_v-r_0]+c.p.\right],
\end{align}
where $c.p.$ means summation over two terms with cyclic permutations of roots $r_0$, $r_1$, and $r_2$.
Here $r_0$, $r_1$, and $r_2$ are functions of $k$ defined in Appendix in Eq.(\ref{eq:rn-roots}) and describe the momentum dispersion of energy bands. The symmetry under permutations of $r_0,r_1$, and
$r_2$ is obvious in these equations. Here we also introduced an ultraviolet cutoff $\Lambda$ for energy, which is of order $\hbar v_F \pi/(a\sqrt{2})$, where $a$ is the lattice constant
$a=2.46 \,\AA$,  and we take $v_F=10^6 \,m/s$ as for graphene \cite{Goerbig}. This cutoff determines the range of applicability of the low-energy model.

\subsection{Properties of gap equations and critical coupling constant}
\label{sec:intravalley-quantitative}

Before solving the gap equations numerically, we should note several their algebraic properties. At
first, if a certain set ${m,\,m_2,\,\mu_v}$ is a solution, then sets with changed signs of masses and valley chemical potential, i.e., ${-m,\,m_2,\,\mu_v}$ and ${m,\,-m_2,\,-\mu_v}$ are also solutions. This follows from the symmetry properties of roots $r_n$ defined in Eq.\eqref{eq:rn-roots}.

Another important property is that there are no solutions of the gap equations (\ref{eq:mu-v-euclid})-(\ref{eq:m2-euclid}) for weak coupling $U$. This can be 
shown in the following way: nontrivial solutions are possible for $U\to0$ only if there are poles in the integrands at $k=0$. This can happen only 
if two bands meet, i.e., $r_{i}(\vec{k}=0)=r_{j}(\vec{k}=0)$. Near the $k=0$ point the denominator is linear in $k$, and the integral over $d^2k$ cancels this singularity. In other words, there are no infrared singularities and therefore nontrivial solutions require a critical value $U_c$ for their appearance.

\begin{figure}
	\centering
	\includegraphics[scale=0.7]{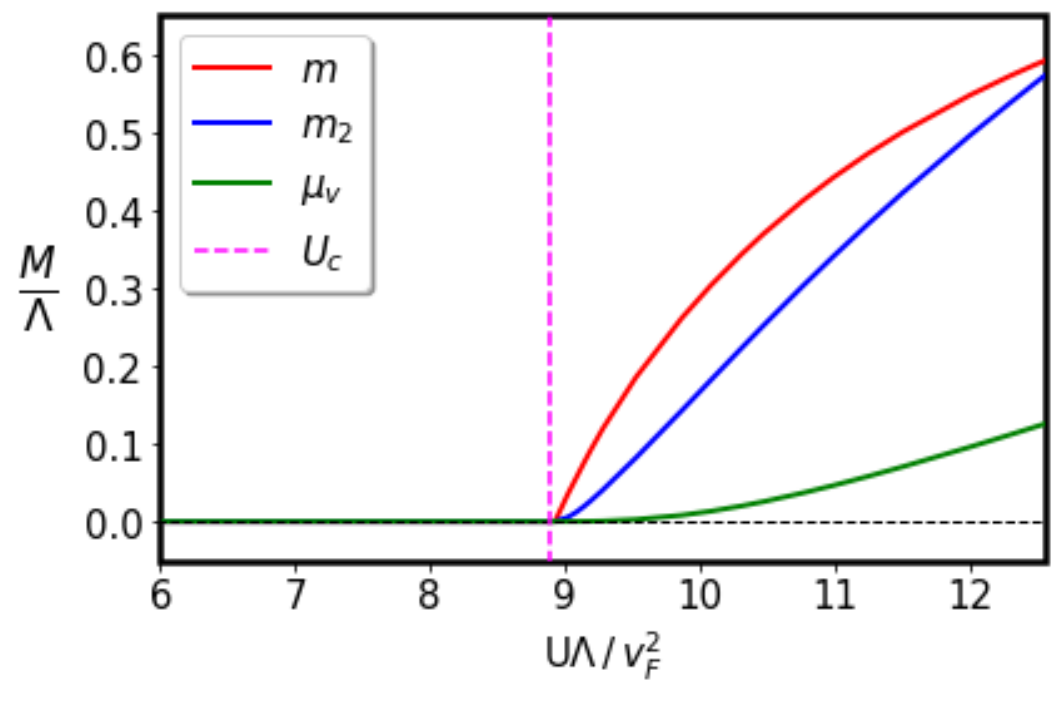}
	\caption{Solutions $M=m,\,m_2, \, \mu_v$ for the system of gap equations \eqref{eq:mu-v-euclid}-\eqref{eq:m2-euclid} as a function of coupling constant $U$. The critical value $U_c$ of coupling constant, estimated in Eq.\eqref{eq:gc-crit-onev}, is marked by dashed vertical line.}
	\label{fig:one-valley-sol}
\end{figure}

Further, let us find the critical coupling constant above which a nontrivial solution exists. Near the critical value, both gaps $m,\,m_2$ and the valley chemical potential $\mu_v$ should tend to zero. Since there are no infrared singularities, the critical coupling constant can be found from the ultraviolet limit of 
the gap equations at large $k$. In such a limit, the gap equation \eqref{eq:m-euclid} reduces to
\begin{align}	m&=m\frac{U}{v_F^2}\int\limits^{\Lambda} \frac{ k dk}{2\pi}\frac{1}{\sqrt{2}k},
\end{align}
which results in the following coupling constant for $m\neq 0$:
\begin{align}\label{eq:gc-crit-onev}
	 U_c=\frac{2\pi\sqrt{2} v_F^2}{\Lambda}\approx 8.89\frac{ v_F^2}{\Lambda}.
\end{align}
Finally, let us proceed to numerical solution of the gap equations. It is convenient to measure $U$ in terms of $v_F^2/\Lambda$. The gap equations \eqref{eq:mu-v-euclid}-\eqref{eq:m2-euclid} form a set of coupled nonlinear equations. We solve them numerically by using standard iterative methods (see, for example, Ref.\cite{Kelley}). Guessing initial points in a wide
range for both masses and valley chemical potential, we were able to find solutions just above the critical constant \eqref{eq:gc-crit-onev}. The corresponding results are shown in Fig.\ref{fig:one-valley-sol}. Near $U=U_c$ gaps $m,\,m_{2}$ are small and valley chemical potential $\mu_v$ is still several orders of magnitude smaller. All these dynamical parameters grow quickly with $U$. We determined also the corresponding critical exponents by using numerically obtained solutions near $U_c$. We found that the dynamical parameters scale as $m\sim (U-U_c)$, $m_2\sim  (U-U_c)^{1.5}$, and $\mu_v\sim (U-U_c)^{3.3}$.

\section{Intervalley gap}
\label{section-intervalley}

Since the denominator in the gap equations (\ref{eq:mu-v-euclid})-(\ref{eq:m2-euclid}) contains the difference of energy dispersions of two bands, this difference
is approximately like that in graphene or two times less. This is the mathematical reason for the existence of a nonzero critical coupling constant for the gap generation like in graphene. However, there is the middle 
completely flat band in the dice model. This suggests that it might be favorable to consider an intervalley gap which couples the electron and holes from different valleys. Additional reason to study such a 
	gap is that similar valley-polarized states are well-known in graphene \cite{Garcia-Pomar,Abergel,Goerbig}. As we will see below, the most crucial property of the intervalley gap is that the difference of the 
energy dispersion of the flat bands in the two valleys does not increase with $k$ at large $k$. The most general two-valley Hamiltonian which describes the intervalley pairing is given by
\begin{align}
H_{2v}=\left(\begin{array}{cc}
H^+_0 & P\\
P^{\dagger} & H^-_0
\end{array}\right),
\label{Hamiltonian-two-valleys}
\end{align}
where we used the short-hand notation $H^{\pm}_{0}=H_{0}(\vec{k},\pm)$
for the free Hamiltonians in the $K$ and $K^{\prime}$ valleys defined by Eq.(\ref{Hamiltonian-free}), and matrix $P$ describes the invervalley gap and, in general, is arbitrary. Since
\be
T H_0^{-}T^{-1}=H_0^{+},\quad T=\left(\begin{array}{ccc}
0 & 0 & 1\\
0 & 1 & 0\\
1 & 0 & 0
\end{array}\right),
\ee
it is convenient to exchange the $A$ and $B$ components of wave functions in the $K^{\prime}$ valley multiplying them by $T$. Then the intervalley Hamiltonian (\ref{Hamiltonian-two-valleys})
takes the form
\begin{align}
H_{iv}=\left(\begin{array}{cc}
H^+_0 & F\\
F^{\dagger} & -H^+_0
\end{array}\right),
\label{Hamiltonian-two-valleys-1}
\end{align}
where its block diagonal elements differ only by sign and $F=PT^{-1}$. Hamiltonian (\ref{Hamiltonian-two-valleys-1}) acts on six-component wave functions
$\psi^T=\left(\psi^K_A,\psi^K_C,\psi^K_B,\psi^{K'}_B,\psi^{K'}_C,\psi^{K'}_A\right)$.
In order to determine the gap equation for the intervalley gap, we need to calculate Green`s function
\begin{equation}
G(\omega,\mathbf{k})=\frac{1}{\omega - H_{iv}}=\left(\begin{array}{cc}
\omega-H^+_0 & F \\
F^{\dagger} & \omega+H^+_0
\end{array}\right)^{-1},
\label{Green-function-two-valleys}
\end{equation}
where $F$ should be determined self-consistently from the Schwinger-Dyson equation which we derive below.

\subsection{Ansatz and gap equation}
\label{section:general-ansatz}

Let us to consider the following ansatz for the intervalley gap with diagonal matrix $F$ whose elements, however, are different:
\begin{equation}
F=\mbox{diag}\,(\Delta,\Delta_2,\Delta)
\label{intervalley-gap-ansatz}
\end{equation}
and, without loss of generality, we assume that $\Delta$ and $\Delta_2$ are real. This specific ansatz, whose first and third diagonal elements are the same, is consistent with the intervalley particle-hole
symmetry (compare it with the particle-hole symmetry (\ref{charge-conjugation}) for the intravalley electron and hole pairing) because the anticommutator of the operator ${\cal C}_{iv}=AKV$ with $H_{iv}$ is zero
\begin{equation}
\left\{{\cal C}_{iv},H_{iv}\right\}=0,\quad V=\left(\begin{array}{cc}
I & 0 \\
0 & -I
\end{array}\right).
\label{charge-conjugation-1}
\end{equation}
Here $A$ is defined in Eq.(\ref{charge-conjugation}), $K$ is the complex conjugation, and $V$ acts on the intervalley indices. The particular form of matrix V is motivated by the order of sublattice wave functions in 6-component spinor and is in agreement with Eq.\eqref{Hamiltonian-two-valleys-1}. Note that since the intervalley particle-hole symmetry is preserved, it is no need
to introduce the valley dependent chemical potential $\xi \mu_v$ like we did in the previous section for the case of intravalley pairing, where $m_2$ breaks the intravalley particle-hole symmetry.
Green`s function (\ref{Green-function-two-valleys}) for the intervalley gap function (\ref{intervalley-gap-ansatz}) is derived in Appendix \ref{GF-inter}.

\begin{figure}
	\centering
\includegraphics[scale=0.65]{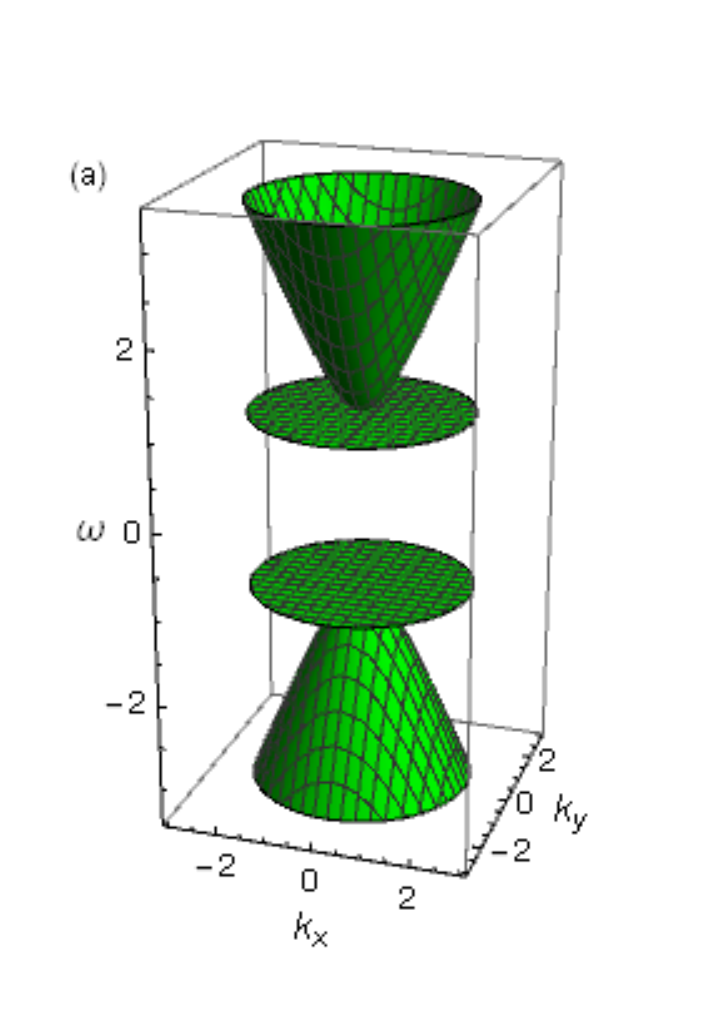}
\includegraphics[scale=0.65]{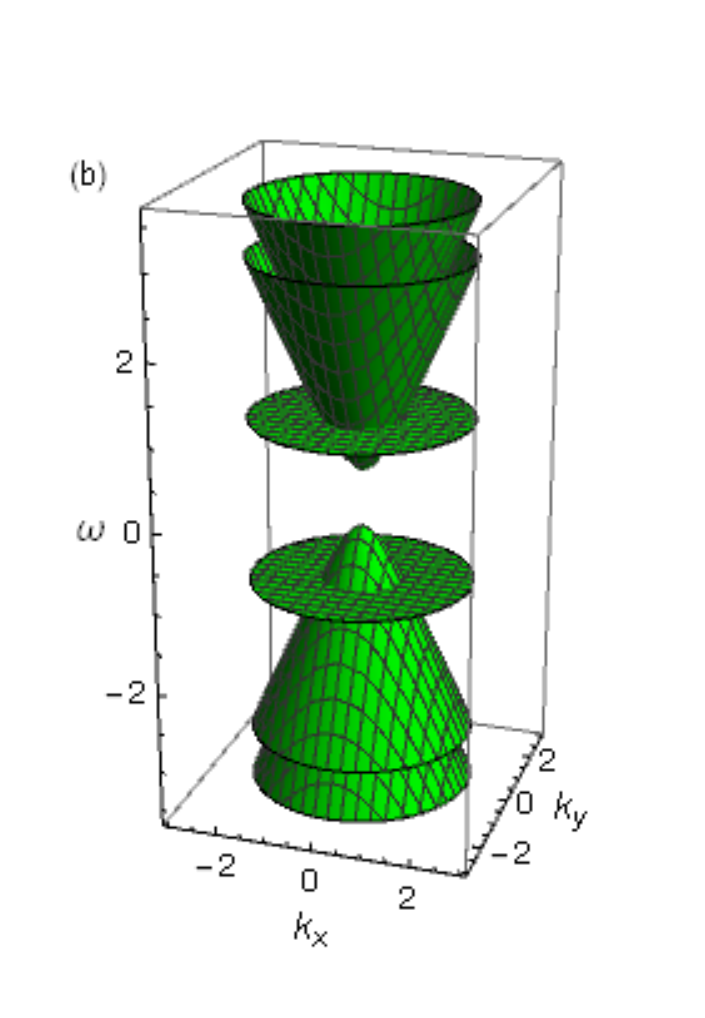}
\includegraphics[scale=0.65]{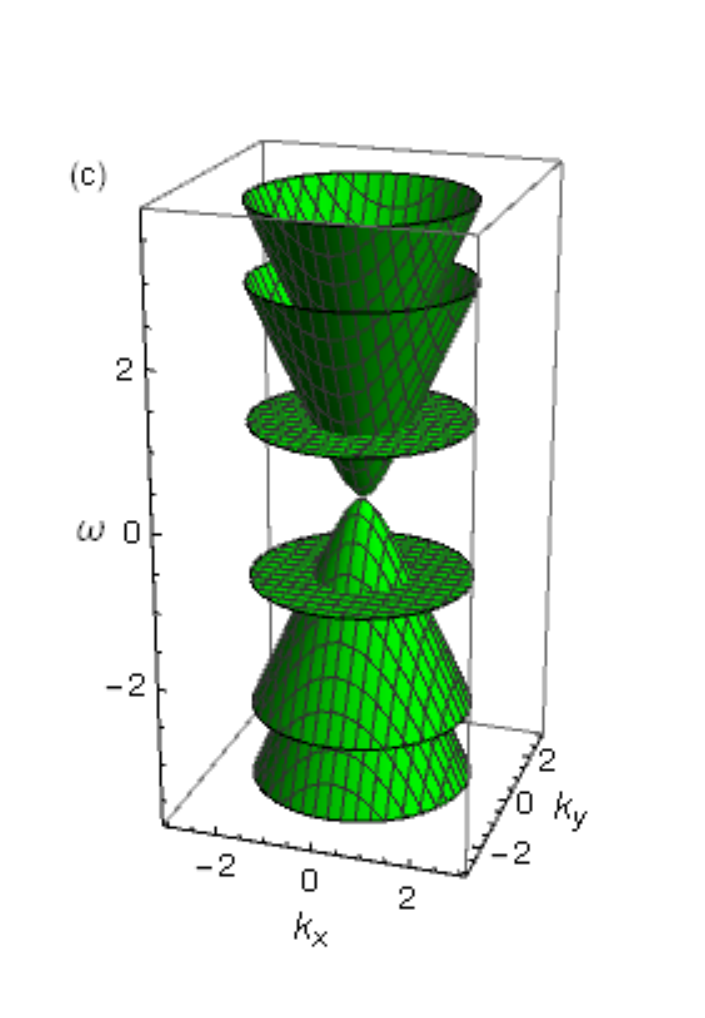}
\caption{Energy dispersion for $\Delta_2=\Delta$ (panel (a)), $\Delta_2=0.35\Delta$ (panel (b)), and $\Delta_2=0.02\Delta$ (panel (c)). Here $\omega$ and $\mathbf{k}$ are measured in units of $\Delta$.}
	\label{fig1:disp-6-6}
\end{figure}

Using this Green's function, we readily find that the Schwinger--Dyson equation leads to the following gap equation:
\begin{align}
\label{eq:2-valley-gap-equation}
	F=i\frac{2U}{v_F^2} \int \frac{d \omega d^{2} k}{(2 \pi)^{3}} \frac{B}{\det[\omega-H_{iv}]},
\end{align}
where $B$ is the off-diagonal block of Green's function defined in Eq.(\ref{element-B}). The determinant in the denominator equals
\begin{equation}
	\det[\omega-H_{iv}]=(\omega^2-\Delta^2)(\omega^2-a^2)(\omega^2-b^2),
	\label{determinant}
\end{equation}
where
\begin{equation}
a^2,b^2=\frac{1}{2}\left(4 k^2+\Delta^2+\Delta^2_2 \pm |\Delta-\Delta_2|\sqrt{8 k^2+(\Delta+\Delta_2)^2}\right).
\label{a-b-coefficients}
\end{equation}
The corresponding spectrum is shown in Fig.\ref{fig1:disp-6-6} for several values of $\Delta$ and $\Delta_2$. We will find below that 
$\Delta_2 \ll \Delta$ for solutions of the gap equations, therefore, panel (c) describes the most relevant case. Equation (\ref{eq:2-valley-gap-equation})
after the Wick rotation $\omega\to i\omega$ gives the equations for gap parameters which can be written as follows:
\begin{align}
\label{gap-equations-intravalley-1}
\Delta &=\frac{2U}{v_F^2} \int \frac{d\omega d^2k}{(2\pi)^3}\left[\frac{A}{\omega^2+a^2}+\frac{B}{\omega^2+b^2}+\frac{C}{\omega^2+\Delta^2}\right],\\
\Delta_2&=\frac{2U}{v_F^2} \int \frac{d\omega d^2k}{(2\pi)^3}\,\left[\frac{\Delta_{2}(a^2 - \Delta ^2)-2 \Delta k^2}{\left(a^2-b^2\right) \left(a^2+\omega^2\right)}+\frac{ \Delta_{2}( \Delta ^2-b^2)+2\Delta  k^2}{\left(a^2-b^2\right)\left(b^2+\omega ^2\right)}\right],
\label{gap-equations-intravalley-2}
\end{align}
where $a^2$ and $b^2$ are defined in Eq.(\ref{a-b-coefficients}) and coefficients $A,B,C$ are
\begin{align}
&A=\frac{a^4\Delta -a^2(\Delta ^3+ \Delta  \Delta_2^2+2 \Delta  k^2+ \Delta _2 k^2)+\Delta(\Delta\Delta _2+k^2)(\Delta\Delta _2+2k^2)}{\left(a^2-b^2\right) \left(a^2-\Delta^2\right)},\nn
&B=A(a\leftrightarrow b),\quad
C=\frac{2 \Delta k^2\left(  k^2-\Delta ^2+\Delta\Delta _2 \right)}{\left(a^2-\Delta ^2\right)\left(b^2-\Delta ^2\right)}.
\end{align}

The structure of the gap equations (\ref{gap-equations-intravalley-1}), (\ref{gap-equations-intravalley-2})  implies that we can assume without loss 
of generality that $\Delta\geq 0$ and leave the sign of $\Delta_{2}$ undefined. Then integrating over frequency and angle, we obtain ($a,b > 0$)
\begin{align}
	\label{inter-Delta-1}
	&\Delta=\frac{2U}{v_F^2} \int_{0}^{\Lambda} \frac{ kdk}{2\pi}\frac{1}{a+b}\left[\frac{k^2(a^2+a \Delta_{2}-k^2)}{a (a-b) (a+\Delta )}+(a\leftrightarrow b)
	+\frac{\Delta}{2}+\frac{\Delta_{2}(\Delta  \Delta_{2}+3k^2)}{2 a b }\right],\\
\label{inter-Delta-2}
&\Delta_2=\frac{2U}{v_F^2}\int_{0}^{\Lambda} \frac{ kdk}{2\pi}\frac{1}{a+b}\left[\frac{\Delta(\Delta\Delta_{2}+2k^2)}{2 a b
	}+\frac{\Delta_{2}}{2 }\right].
\end{align}
The above equations form a coupled system of equations for $\Delta$ and $\Delta_2$. Note the symmetry under the exchange $a\leftrightarrow b$. We will solve this system numerically in
Subsec.\ref{intervalley-numerical}. As we argued above, the flat band should play the principal role for intervalley gap generation. Therefore, before finding numerical solutions to the gap
equations \eqref{inter-Delta-1} and (\ref{inter-Delta-2}), it is instructive to study in the next subsection the intervalley gap generation by retaining only
the flat bands in the two valleys.

\subsection{Flat band approximation}
\label{section:FBA}

To study the intervalley gap generation in the flat band approximation (FBA), we should find explicitly the corresponding flat band electron states. First of all, by using Eq.(\ref{Hamiltonian-free}), we obtain
that the normalized states of zero energy of the free Hamiltonian $H^+_0$ are given by
\begin{equation}
\psi^T_0(\mathbf{k})=\frac{1}{\sqrt{2}\,2\pi}\left(1,\,0,\,-\frac{k_+}{k_-}\right),
\label{flat-band-states}
\end{equation}
where $k_{\pm}=k_{x} \pm i k_{y}$. In order to proceed and consider the intervalley gap generation, we should determine the eigenstates of Hamiltonian (\ref{Hamiltonian-two-valleys-1}) in the
subspace composed of flat band states in two valleys, i.e.,
\begin{equation}
H_{iv}\Psi=E\Psi,
\label{eigenstate-equation}
\end{equation}
where $\Psi$ consists of the flat band states (\ref{flat-band-states}) in two valleys with two unknown constants $C_1 \equiv N$ and $C_2 \equiv NC$
\be
\Psi^T=N\,\left(1,\,0,\,-\frac{k_+}{k_-},\,C,\,0,\,-C\frac{k_+}{k_-}\right).
\ee
The eigenstate equation (\ref{eigenstate-equation}) for $F=\mbox{diag}\,(\Delta,\Delta_2,\Delta)$ gives two nontrivial relations
\begin{equation}
E-\Delta C=0,\quad \Delta - E C=0.
\label{system-equations}
\end{equation}
Note that the gap $\Delta_2$ is not present in the above equations. Clearly, the system of equations (\ref{system-equations}) means that there are
two solutions
\begin{equation}
C=1,\quad E=-\Delta,\quad\quad\quad\quad\quad C=-1,\quad E=\Delta.
\label{fba-solutions}
\end{equation}
Obviously, the two former degenerate flat band solutions in two valleys are now split in energy by $2\Delta$.

Green`s function connected with the flat band states has the form
\begin{equation}
G_{FB}(\omega,\mathbf{k})=\frac{\Psi_{-\Delta}\Psi^{\dagger}_{-\Delta}}{\omega+\Delta} + \frac{\Psi_{\Delta}\Psi^{\dagger}_{\Delta}}{\omega-\Delta},
\label{FBA-Green-function}
\end{equation}
where
\be
\Psi^T_{-\Delta}= \frac{1}{4\pi}\,\left(1,\,0,\,-\frac{k_+}{k_-},\,-1,\,0,\,\frac{k_+}{k_-}\right),\quad
\Psi^T_{\Delta}= \frac{1}{4\pi}\,\left(1,\,0,\,-\frac{k_+}{k_-},\,1,\,0,\,-\frac{k_+}{k_-}\right).
\ee
In order to study the gap generation, we should consider the off-diagonal elements of the matrix $G_{FB}$. Let us consider the upper off-diagonal block (the consideration of the lower off-diagonal block gives the same results). Since the element $25$ of the matrix $G_{FB}$ is zero, we conclude that $\Delta_2=0$ in
the flat band approximation. The elements $14$ and $36$ of the matrix $G_{FB}$ coincide. Therefore, the ansatz with $F=\mbox{diag}\,(\Delta,\Delta_2,\Delta)$,
whose $11$ and $33$ elements are the same, is indeed consistent. Thus, we have the following gap equation in the flat band approximation defined by the element $14$
or $36$ of the matrix $G_{FB}$:
\begin{equation}
\Delta=-i\frac{2U}{v_F^2} \int \frac{d\omega d^2k}{(2\pi)^3}\,\frac{1}{4\pi}\,\left(\,-\frac{1}{\omega-\Delta} + \frac{1}{\omega+\Delta}\,\right)=\frac{iU}{\pi v_F^2} \int \frac{d\omega d^2k}{(2\pi)^3}\,\frac{\Delta}{\omega^2-\Delta^2
+i\delta}.
\label{FBA-gap-equation}
\end{equation}
Integrating over $\omega$ and introducing a cut-off $\Lambda$ over momentum, we easily find the following gap in the flat band approximation:
\begin{equation}
\Delta=\frac{U\Lambda^2}{8\pi^2 v_F^2}.
\label{FBA-gap}
\end{equation}
Clearly, the critical coupling constant is zero like in the case of the magnetic catalysis of the gap generation [\onlinecite{GMSh}] in a model with local four-fermion interaction in $2+1$ dimensions. Note that there is no trivial solution again as in the magnetic catalysis case. The calculated gap (\ref{FBA-gap}) is quadratically divergent and is much larger than the gap in the lowest Landau level (LLL) approximation. The reason is that Green`s function in the LLL approximation in fermion systems with relativistic-like energy dispersion and dynamically generated mass $m$ is quite similar to the flat band Green`s function (\ref{FBA-Green-function})
\be
S_{LLL}(q)=e^{-\frac{\mathbf{q}^2}{|eB|}}\,\frac{\omega\gamma_0-m}{\omega^2-m^2}\,(1-i\gamma_1\gamma_2)
\ee
except that it contains an exponentially decreasing factor in momenta $\mathbf{q}$ (here $\gamma_0$, $\gamma_1$, and $\gamma_2$ are the Dirac $\gamma$-matrices). Therefore, the corresponding solution to the gap 
equation is proportional to the magnetic field strength $|eB|$ rather than the cut-off squared $\Lambda^2$ (see Eq.(64) in [\onlinecite{GMSh}]). This is the reason why the intravalley gap is so large.

\begin{figure}
	\centering
\includegraphics[scale=0.7]{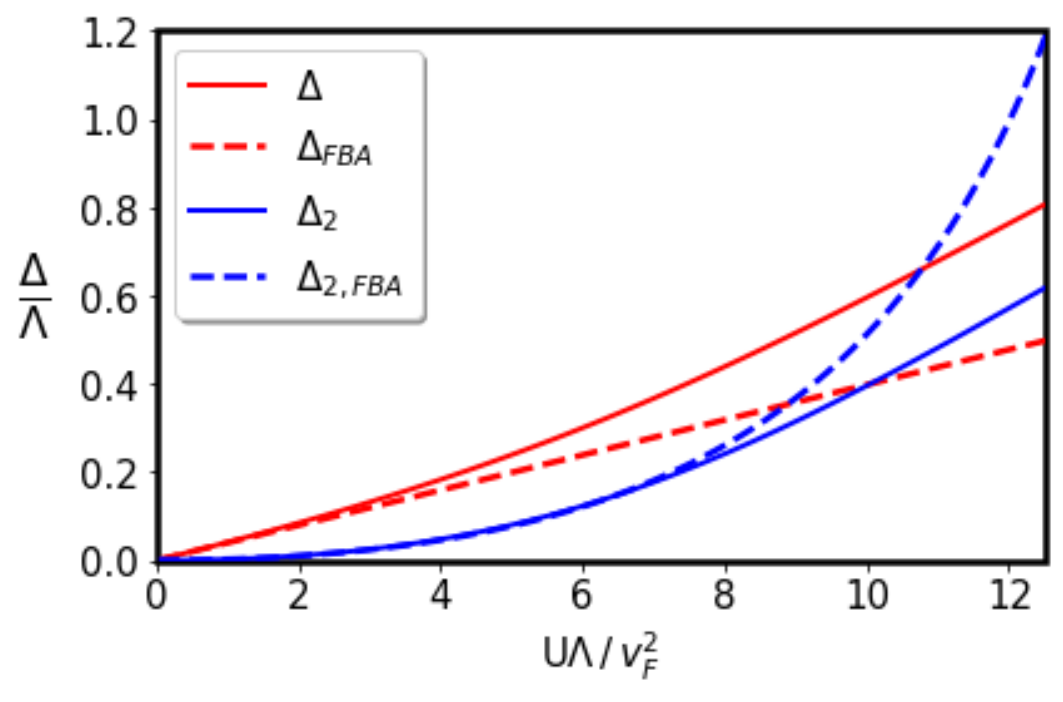}
\caption{Numerical solutions of the gap equations \eqref{inter-Delta-1} and \eqref{inter-Delta-2}. }
	\label{fig:intervalley-gap-solutions}
\end{figure}

In addition, we should note that the flat band approximation in the model under consideration can be obtained as a large momentum limit of gap equations. Assuming that $\Delta,\,\Delta_{2}\ll\Lambda$, we can
approximate coefficients $a,b$ in (\ref{a-b-coefficients}) as follows:
\begin{align}
	&a^{2},\,b^2
\approx 2k^2.
\end{align}
Substituting this back in Eqs.(\ref{inter-Delta-1}) and \eqref{inter-Delta-2}, we find the following solutions for gap parameters:
\begin{align}\label{eq:Delta-2-FBA-extended}
\Delta=\frac{U}{v_F^2}\frac{\Lambda^2}{8\pi},\quad 	\Delta_{2}=\frac{\sqrt{2}U\Lambda\Delta}{8\pi v_F^2-\sqrt{2}U\Lambda}=\frac{\sqrt{2}}{1-\sqrt{2}\Delta/\Lambda}
\frac{\Delta^2}{\Lambda}.
\end{align}
These expressions extend the results obtained in the two-band FBA discussed above and incorporate corrections from other energy bands for $\Delta_2$. Before proceeding to the numerical analysis, it is
instructive to estimate the values of generated gaps. Using cut-off $\Lambda=v_{F} \pi /(a \sqrt{2})$, 
we find $\Delta=\pi U /(16 a^2)$. For local Coulomb interaction, we can use the corresponding estimate in graphene $V_C=e^2\sqrt{3}/(a\pi) \approx 3.3\,\text{eV}$ \cite{Herbut2006}. This gives the coupling constant
$U=V_C/\Omega_{BZ}$ (here $\Omega_{BZ}=2/(\sqrt{3}a^2)$ is the area of the Brillouin zone), we find
$\Delta= 0.56 \,\text{eV}$. Interestingly, the obtained result qualitatively agrees with the study of gap generation in twisted graphene
bilayers near a magic angle \cite{Cea2020PRB}, where the flat band is present. Indeed, due to the very large length of the moire lattice unit $a_{TBG} \approx 12\,nm$, the corresponding gaps are suppressed by factor $a^2/a^2_{TBG}$ leading to gaps of order few
$meV$ in twisted bilayer graphene.
Finally, we note that it is crucial that there are two flat bands in different valleys and our analysis shows that the presence of a single flat band is not sufficient for the gap generation for an arbitrary small coupling constant.

\subsection{Numerical analysis of solutions and their free energy}
\label{intervalley-numerical}

In the numerical analysis, it is convenient to measure $U$ in units of $v^2_F/\Lambda$. Like in Sec.\ref{sec:intravalley-quantitative} we use the iteration method to solve the gap equations. The corresponding 
numerical solutions are presented in Fig.\ref{fig:intervalley-gap-solutions} and are compared with the flat band approximation result.

One should note that gap $\Delta_2$ is one order of magnitude smaller that gap $\Delta$ for small values of $U$ such that $U\Lambda/v^2_F<2$. For example, at $U\Lambda/v^2_F=1.4$ we find $\Delta\approx 0.06 \Lambda,\,\,\Delta_{2} \approx 0.005 \Lambda$. However, $\Delta_{2}$ grows much faster with coupling constant $U$, approximately as
$U^2$, which quantitatively agrees with Eq.\eqref{eq:Delta-2-FBA-extended} at small coupling constant. $U\Lambda/v^2_F >4$, the FBA solution starts to deviate from the exact solution.
Of course, we should note that the low-energy model is not applicable when gaps become of order $\Lambda$.

\begin{figure}
	\centering
\includegraphics[scale=0.65]{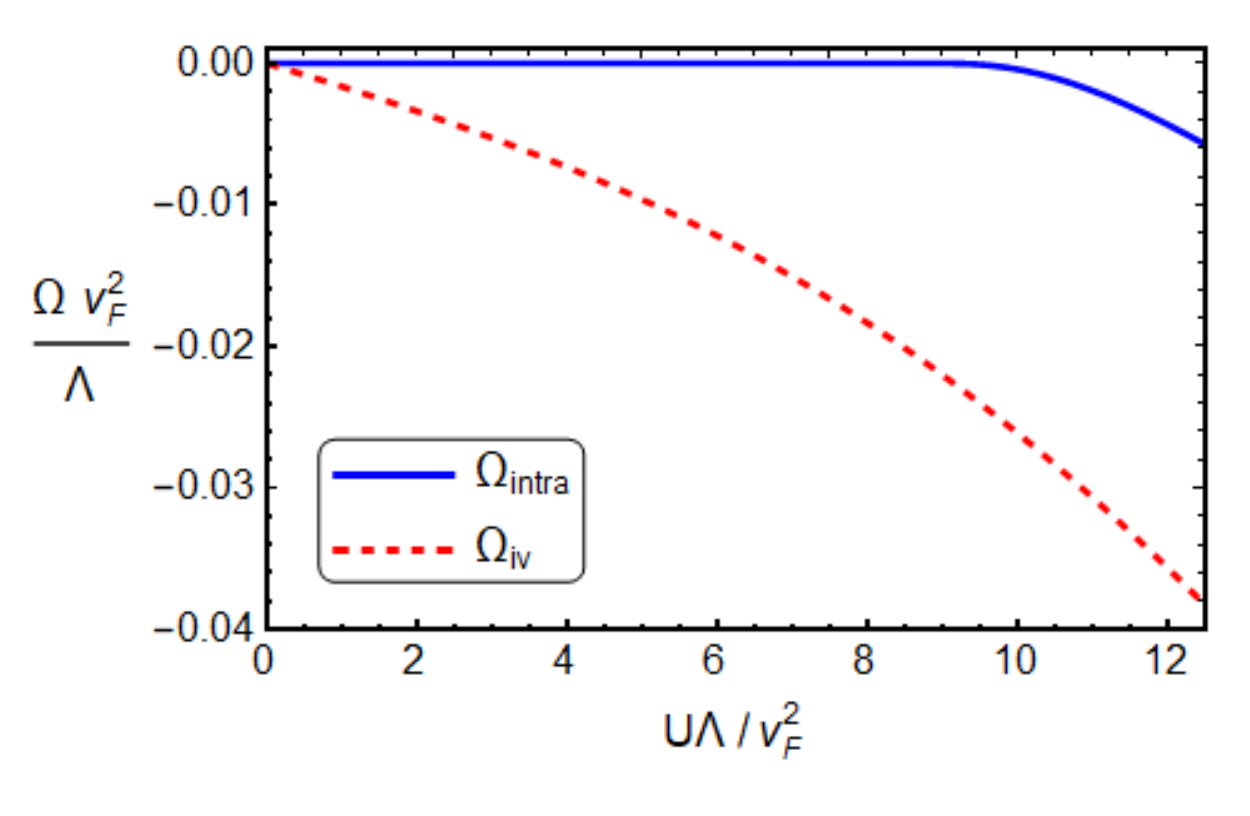}
	\caption{Numerical results for the free energy density $\Omega$ as a function of coupling constant $U$. The free energy density $\Omega_{intra}$ for the intravalley gap solution is given by Eq.\eqref{eq:gamma-intravalley} and $\Omega_{iv}$ for the 
	intervalley gap solution is defined
	in Eq.\eqref{eq:gamma-intervalley}.}
	\label{fig:free-energy}
\end{figure}

Among all solutions of the Schwinger Dyson equation the stable one is selected as the solution with the lowest free-energy density.
The free energy density of a certain solution is determined by the value of the Baym--Kadanoff effective action (\ref{BK-action}) for the corresponding extremum
of the Schwinger-Dyson equation $\delta\Gamma(G)/\delta G =0$ which takes the form \cite{Gorbar2008PRB}
\begin{align}
	\Gamma=-\mathrm{i} \operatorname{Tr}\left[\operatorname{Ln} G^{-1}S+\frac{1}{2}\left(S^{-1} G-1\right)\right].
\label{free-energy}
\end{align}
The free energy density is given by $\Omega=-\Gamma / T V$ where $TV$ is a space-time volume. Integrating by parts the logarithm term and
omitting the irrelevant surface term (which does not depend on gaps), we find
\begin{align}\label{eq:gamma-normalized}
	\Omega=&i\int\limits_{-\infty}^{\infty} \frac{d \omega}{2 \pi} \frac{2}{v_F^2}\int\frac{d^2k}{(2\pi)^2}\tr\left\{-\omega\left[ \frac{\partial G^{-1}(\omega)}{\partial \omega} G(\omega)+S^{-1}(\omega)\frac{\partial S(\omega)}{\partial \omega}\right]+\frac{1}{2}\left[S^{-1}(\omega) G(\omega)-1\right]\right\}.
\end{align}
The technical details of calculation of the energy density of the intravalley and intervalley gap solutions are presented in Appendix \ref{appendix:free-energy}. Here we present the results of numerical evaluation
by using Eqs.\eqref{eq:gamma-intravalley} and \eqref{eq:gamma-intervalley} and plot the free energies for both types of gaps in Fig.\ref{fig:free-energy}. Clearly, the intervalley gap solution is always preferable including the region above the critical coupling constant \eqref{eq:gc-crit-onev}.

\section{Summary}
\label{sec:summary}

We studied the gap generation in the dice model at the neutrality point. We found that there are two main intravalley and intervalley
types of the electron-hole pairing which pairs the electron and hole states in the same and different valleys, respectively. The neutrality of the system provides an important reduction of the number of order parameters. Indeed, it turned out that the particle-hole symmetry restricts the number of possible order parameters
to three in the case of the intravalley gap and the intervalley particle-hole symmetry gives two independent order parameters for the intervalley pairing. Thus, there
are three and two gap equations in the case of the intra- and intervalley gap generation, respectively.

To get an insight into the gap generation in the dice model and reveal the role of the flat band, we employed a local four-fermion interaction in our study. The
main technical advantage of local interaction is that the gap equations are algebraic and admit an efficient numerical and partially analytic analysis. Our main finding is that the intervalley gap is generated for an arbitrary small coupling constant unlike the intravalley gap which requires a critical coupling constant. These qualitatively different results are due to the crucial role which plays the flat band in the intra- and intervalley gap generation.

Indeed, the intravalley gap pairs the electron and hole states in the same valley, therefore, it cannot pair states from the flat band only because such states cannot be the electron and hole ones simultaneously. In contrast, the intervalley gap relates the electron and hole states in flat bands from different valleys. The dispersionless band has a singular density of states that strongly enhances the intervalley gap generation leading to an extremely large gap proportional to the coupling constant times the area of the Brillouin zone. This result agrees with the heuristic argument that the completely flat band is the most favorable for the gap generation \cite{Khodel1990,Volovik1991,Volovik1994,Volovik2019}.
The intervalley gap generation in the dice model is also qualitatively similar to that in the case of magnetic catalysis in (2+1) dimensions in fermion systems with relativistic-like energy spectrum [\onlinecite{GMSh}]. Indeed, magnetic field produces completely flat Landau levels and a fermion gap is generated for an arbitrary small coupling constant and is proportional to the degeneracy of the lowest Landau level defined by the inverse of the magnetic length squared $l^{-2}$. Since the magnetic length is typically much larger than the lattice constant (e.g., in graphene, $l$ is 26 $nm$ at $B=1\, T$ and the lattice constant $a=0.246\,nm$), the intravalley gap is strongly enhanced in the dice model by factor $l^2/a^2 \approx 10^4$ compared to the gap generated due to the magnetic catalysis. Thus, we conclude that the flat band catalysis is very efficient indeed. The underlying physical reason is very simple. Due to the dispersionless flat band, the integration over momentum leads to a gap proportional to the area of the Brillouin zone, thus, very strongly enhancing the gap. This means that even if the middle band is not completely flat, still the intervalley gap generation should be very efficient and robust. Finally, we note that our results emphasize and shed additional light on the important role of flat band in the gap generation for magic angle twisted bilayer graphene.

\begin{acknowledgements}
E.V.G. and V.P.G. acknowledge collaboration within the Ukrainian-Israeli Scientific Research Program of the Ministry of Education and Science of Ukraine
(MESU) and the Ministry of Science and Technology of the state of Israel (MOST).
\end{acknowledgements}

\appendix

\section{Intravalley Green`s function and gap equations}
\label{GF-intra}

Green`s function of quasiparticles in the dice model with intravalley gaps at given valley $\xi$ in momentum space equals
\begin{eqnarray}
&&G_{\xi}(\omega,\mathbf{k})=\frac{1}{\omega- H_{\xi}+\xi \mu_v} = \frac{1}{\det[\omega-H_{\xi}+\xi\mu_v]} \nonumber\\
&&\times
\left(\begin{array}{ccc}
(\omega+\xi \mu_v+m)(\omega+\xi \mu_v-\xi m_2)-k^2 & D & E\\
B & (\omega+\xi \mu_v)^2-m^2 & H\\
C & F & (\omega+\xi \mu_v-m)(\omega+\xi \mu_v-\xi m_2)-k^2
\end{array}\right),\\
&&\det[\omega-H_{\xi}+\xi\mu_v]=(\omega+\xi \mu_v-\xi m_2)((\omega+\xi \mu_v)^2-m^2)-2\tilde{k}^2(\omega+\xi \mu_v)\nonumber,
\label{Green-function}
\end{eqnarray}
where off-diagonal matrix elements are
\begin{eqnarray}
&&D=(\omega+\xi \mu_v+m)k^{\xi}_-,\quad E=(k^{\xi}_-)^2,\quad B=(\omega+\xi \mu_v+m)k^{\xi}_+,\nonumber\\
&& H=(\omega+\xi \mu_v-m)k^{\xi}_-,\quad C=(k^{\xi}_+)^2,\quad F=(\omega+\xi \mu_v-m)k^{\xi}_+,
\end{eqnarray}
and $k^{\xi}_-=(\xi k_x-ik_y)/\sqrt{2}$ and $k^{\xi}_+=(\xi k_x+ik_y)/\sqrt{2}$.

Clearly, all off-diagonal terms in $G_{\xi}(\omega,\mathbf{k})$ depend linearly or quadratically on $k^{\xi}_+$ and $k^{\xi}_-$, therefore, all such terms vanish after integration over momentum in Eq.(\ref{SD-equation-intra}). Hence the Schwinger--Dyson equation gives three equations
for $\mu_v$, $m$, and $m_2$ for the diagonal terms. They are
\begin{align}
\label{equation-m-v}
&\xi \mu_v=-i\frac{2U}{v_F^2}\int \frac{d\omega d^2k}{(2\pi)^3}\,\,\frac{(\omega+\xi \mu_v)(\omega+\xi \mu_v -\xi m_2)-k^2}{\det[\omega+i0\mbox{sgn}(\omega)-H_{\xi}+\xi\mu_v]},\\
\label{equation-m}
&m=i\frac{2U}{v_F^2}\int \frac{d\omega d^2k}{(2\pi)^3}\,\,\frac{m(\omega+\xi \mu_v-\xi m_2)}{\det[\omega+i0\mbox{sgn}(\omega)-H_{\xi}+\xi\mu_v]},\\
\label{equation-m2}
&\xi m_2=i\frac{2U}{v_F^2}\int \frac{d\omega d^2k}{(2\pi)^3}\,\,\frac{(\omega+\xi \mu_v)\xi m_2+k^2-m^2}{\det[\omega+i0\mbox{sgn}(\omega)-H_{\xi}+\xi\mu_v]}.
\end{align}
Note that Eq.(\ref{equation-m}) for gap $m$ is explicitly homogeneous unlike Eqs.(\ref{equation-m-v}) and (\ref{equation-m2}) for $\mu_v$ and $m_2$. As we stated above, we seek solutions with $m \ne 0$, otherwise,
the flat band with $\epsilon=0$ is realized and it is not clear how to define a half-filled state.

Since $\xi$ equals $\pm$ in two valleys, in fact, the system of equations (\ref{equation-m-v})-(\ref{equation-m2}) consists of six equations for three unknowns $\mu_v$, $m$, and $m_2$. It is convenient to change the variable $\omega \to \xi\omega$ on the right-hand side of these equations to see that this system of equations is consistent. In order to calculate the integral over $\omega$ in the above gap equations and make it explicitly convergent we
represent the integrands as $I(\omega)=[I(\omega) + I(-\omega)]/2$ utilizing the symmetric integration in $\omega$. The denominators in the integrands is convenient to write in terms of roots of the cubic equation
\begin{equation}
\det[\omega-H_{\xi}+\xi\mu_v]=(\omega+\xi\mu_v-m_2)((\omega+\xi\mu_v)^2-m^2)-2 k^2(\omega+\xi\mu_v)=0
\label{roots}
\end{equation}
which are given by
\begin{align}\label{eq:rn-roots}
	r_n=\omega_n+\mu_v=\frac{m_2}{3} + 2m\sqrt{-\frac{p}{3}}\,\cos\left(\frac{1}{3}\,\arccos\,\left(\frac{3q}{2p}\sqrt{-\frac{3}{p}}\right)-\frac{2\pi n}{3}\right),\quad\quad n=0,1,2,
\end{align}
where
\begin{equation}
p=-\left(1+\frac{2 k^2}{m^2}+\frac{m^2_2}{3m^2}\right),\quad\quad q=\frac{m_2}{m}\left(1-\frac{1+\frac{2 k^2}{m^2}}{3} -\frac{2m^2_2}{27m^2}\right)= \frac{m_2}{m}\left(\frac{2}{3}\left(1-\frac{k^2}{m^2}\right) -\frac{2m^2_2}{27m^2}\right).
\label{pq-definition}
\end{equation}
Thus, the determinant can be conveniently rewritten as
\begin{align}
	\det[\omega-H_{\xi}+\xi\mu_v]=(\omega+\xi\mu_v-r_0)(\omega+\xi\mu_v-r_1)(\omega+\xi\mu_v-r_2).
\end{align}
Then we obtain
\begin{eqnarray}
\label{equation-m-v-2}
&&\mu_v=-\frac{iU}{v_F^2}\int \frac{d\omega d^2k}{(2\pi)^3}\,\left(\frac{(\omega+\mu_v)(\omega+\mu_v -m_2)-k^2}
{\det[\omega+i\delta-H_{\xi}+\xi\mu_v]}  -\left[ \mu_v \to -\mu_v , m_2 \to - m_2 \right]\right),\\
\label{equation-m-2}
&&m=\frac{iU}{v_F^2}\int \frac{d\omega d^2k}{(2\pi)^3}\,\left(\frac{m(\omega+\mu_v-m_2)}{\det[\omega+i\delta-H_{\xi}+\xi\mu_v]}  + \left[ \mu_v \to -\mu_v , m_2 \to - m_2 \right]\right),\\
&&m_2=\frac{iU}{v_F^2}\int \frac{d\omega d^2k}{(2\pi)^3}\,\left(\frac{(\omega+\mu_v)m_2+k^2-m^2}{\det[\omega+i\delta-H_{\xi}+\xi\mu_v]}- \left[ \mu_v\to -\mu_v , m_2 \to - m_2 \right]\right),
\label{equation-m2-2}
\end{eqnarray}
where $\delta=0\mbox{sgn}(\omega)$. This form of equations is convenient for further integration over frequency leading to
Eqs.(\ref{eq:mu-v-euclid}) - (\ref{eq:m2-euclid}) in the main text.

\section{Intervalley Green`s function}
\label{GF-inter}

For Green`s function of the intervalley gap ansatz (\ref{Green-function-two-valleys}), we find the following explicit expression:
\begin{equation}\label{eq:green-function}
	G_{ij}=\frac{1}{\det[\omega-H_{iv}]}\,\left(\begin{array}{cc}
A & B \\
C & D
\end{array}\right),\quad\quad \det[\omega-H_{iv}]=(\omega^2-\Delta^2)\left[\omega^4-\omega^2(4 k^4+\Delta^2+\Delta^2_2) + (2 k^2+\Delta\Delta_2)^2 \right].
\end{equation}
The elements of the matrix $A$ are
\begin{align}
&A_{11}=
	\omega  \left(\left(\Delta ^2-\omega ^2\right) \left(\Delta _2^2-\omega ^2\right)+2 k^4+\tilde{k}^2 \left(\Delta ^2+2 \Delta _2
	\Delta -3 \omega ^2\right)\right),\quad
	&A_{12}=k_- \left(\Delta ^2-\omega ^2\right) \left(\Delta  \Delta _2+2 k^2-\omega ^2\right),\nn
&A_{13}=k^2_- \omega  \left(\Delta ^2-2 \Delta  \Delta _2-2 k^2+\omega ^2\right),\quad	
	&A_{21}=k_+ \left(\Delta ^2-\omega ^2\right) \left(\Delta  \Delta _2+2 k^2-\omega ^2\right),\nn
&A_{22}=\omega
	\left(\Delta ^2-\omega ^2\right) \left(\Delta ^2+2 k^2-\omega ^2\right),\quad
&A_{23}=k_- \left(\Delta ^2-\omega ^2\right)
	\left(\Delta  \Delta _2+2 k^2-\omega ^2\right),\nn
&A_{31}=k^2_+ \omega  \left(\Delta ^2-2 \Delta  \Delta _2-2 k^2+\omega ^2\right),\quad
&A_{32}=k_+ \left(\Delta
	^2-\omega ^2\right) \left(\Delta  \Delta _2+2 k^2-\omega ^2\right),\nn
&A_{33}=\omega  \left(\left(\Delta
	^2-\omega ^2\right) \left(\Delta _2^2-\omega ^2\right)+2 k^4+k^2 \left(\Delta ^2+2 \Delta _2 \Delta -3 \omega
	^2\right)\right).&
\end{align}
It turned out that $B=C$ and the elements of $B$ are
\begin{align}
	&B_{11}=
	\Delta  \left(\Delta ^2-\omega ^2\right) \left(\Delta _2^2-\omega ^2\right)+2 \Delta  k^4+k^2 \left(3 \Delta ^2 \Delta
	_2-\left(2 \Delta +\Delta _2\right) \omega ^2\right),
& \nn
&B_{12}=\left(\Delta -\Delta _2\right) k_- \omega  \left(\Delta ^2-\omega
	^2\right),\quad
	&B_{13}=-k_{-}^2 \left(\Delta ^2 \Delta _2-2 \Delta  \omega ^2+\Delta _2 \omega ^2+2 \Delta
	k^2\right),
\nn
&B_{21}=\left(\Delta -\Delta _2\right) (-k_+) \omega  \left(\Delta ^2-\omega ^2\right),\quad
&B_{22}=\left(\Delta
	^2-\omega ^2\right) \left(\Delta _2 \left(\Delta ^2-\omega ^2\right)+2 \Delta  k^2\right),\nn
&B_{23}=\left(\Delta -\Delta
	_2\right) (-k_-) \omega  \left(\Delta ^2-\omega ^2\right),\quad
	&B_{31}=-k^2_+ \left(\Delta ^2 \Delta _2-2 \Delta  \omega ^2+\Delta _2 \omega ^2+2 \Delta  k^2\right),\nn
	&B_{32}=\left(\Delta -\Delta _2\right) k_+ \omega  \left(\Delta ^2-\omega ^2\right),&\nn
	&B_{33}=\Delta  \left(\Delta
	^2-\omega ^2\right) \left(\Delta _2^2-\omega ^2\right)+2 \Delta  k^4+k^2 \left(3 \Delta ^2 \Delta _2-\left(2 \Delta
	+\Delta _2\right) \omega ^2\right).&
\label{element-B}
\end{align}
Finally, the elements of $D$ are
	\begin{align}
		&D_{11}=
		\omega  \left(\left(\Delta ^2-\omega ^2\right) \left(\Delta _2^2-\omega ^2\right)+2 k^4+k^2 \left(\Delta ^2+2 \Delta _2
		\Delta -3 \omega ^2\right)\right),&\nn
		&D_{22}=\omega
		\left(\Delta ^2-\omega ^2\right) \left(\Delta ^2+2 k^2-\omega ^2\right),\quad&\nn
		&D_{33}=\omega  \left(\left(\Delta
		^2-\omega ^2\right) \left(\Delta _2^2-\omega ^2\right)+2 k^4+k^2 \left(\Delta ^2+2 \Delta _2 \Delta -3 \omega
		^2\right)\right).& \nn
		&D_{21}=-k_{+}(\Delta^2-\omega^2)(2k^2-\omega^2+\Delta\Delta_2),&D_{31}=k_{+}^2\omega(-2k^2+\Delta^2+\omega^2-2\Delta\Delta_2),\nn
		&D_{12}=-k_{-}(\Delta^2-\omega^2)(2k^2-\omega^2+\Delta\Delta_2),&D_{32}=-k_{+}(\Delta^2-\omega^2)(2k^2-\omega^2+\Delta\Delta_2),\nn
		&D_{13}=k_{-}^2\omega(-2k^2+\Delta^2+\omega^2-2\Delta\Delta_2),&D_{23}=-k_{-}(\Delta^2-\omega^2)(2k^2-\omega^2+\Delta\Delta_2)
	\end{align}
In the main text we use the diagonal elements of $B$ to write the gap equations in the explicit form. Note, that off-diagonal components vanish after integration over polar angle $\phi$ in momentum space.

\section{Evaluation of free energy}
\label{appendix:free-energy}

In this Appendix we present the detailed calculation of the free energy density for intravalley and intervalley gap states in the dice model. The final results
are given by Eqs.\eqref{eq:gamma-intravalley} and \eqref{eq:gamma-intervalley}.


Using expression \eqref{eq:gamma-normalized} for the Baym--Kadanoff free energy, we denote the integrand as
\begin{align}
	&\tilde{\Omega}(\vec{k},\omega)=\tr\left\{-\omega\left[ \frac{\partial G^{-1}(\omega)}{\partial \omega} G(\omega)+S^{-1}(\omega)\frac{\partial S(\omega)}{\partial \omega}\right]+\frac{1}{2}\left[S^{-1}(\omega) G(\omega)-1\right]\right\}.
\label{BK-action-integrand}
\end{align}
First we evaluate the trace and perform summation over valleys, decomposing the result into fractions.
Next it is convenient to perform the Wick rotation $\omega\to i\omega$. For the intravalley gap state, we obtain
\begin{align}
	&\tilde{\Omega}_{intra}(\vec{k},i\omega)=-\frac{8k^2}{\omega^2+2k^2}-\left(\frac{\left(\mu _v-r_0\right) \left(-2 k^2 \left(4 r_0+\mu _v\right)+m^2 \left(3 m_2-2 r_0-\mu _v\right)+r_0 \left(3 r_0 \mu
		_v-m_2 \left(r_0+2 \mu _v\right)\right)\right)}{\left(r_1-r_0\right) \left(r_0-r_2\right) \left(\left(\mu
		_v-r_0\right){}^2+\omega ^2\right)}+c.p.\right),
\end{align}
where $(c.p.)$ denotes cyclic permutation of $r_i$. The integration over frequency $\omega$ is easily performed and
we come at the free energy density for the intravalley gap state given by
\begin{align}\label{eq:gamma-intravalley}
	&\Omega_{intra}=\frac{1}{v_F^2}\int\limits_{0}^{\Lambda}\frac{k\, dk}{\pi}\left[2\sqrt{2}k+\left(\frac{-2 k^2 \left(4 r_0+\mu _v\right)+m^2 \left(3 m_2-2 r_0-\mu _v\right)+r_0 \left(3 r_0 \mu
		_v-m_2 \left(r_0+2 \mu _v\right)\right)}{2\left(r_1-r_0\right) \left(r_0-r_2\right) }\sign[\mu_v-r_0]+c.p.\right)\right],
\end{align}
This expression is invariant under the change of sign $m\to-m$ or $(m_2,\mu_v)\to (-m_2,-\mu_v)$. Using the numerically found solutions from Sec.\ref{sec:intravalley} (see Fig.\ref{fig:one-valley-sol}), we
evaluate the integral over $k$. The corresponding results for the free energy are shown in Fig.\ref{fig:free-energy}.


In the case of the intervalley gap state, we obtain for the integrand in the Baym--Kadanoff free energy \eqref{eq:gamma-normalized} after the Wick rotation $\omega\to i\omega$
\begin{align}
	\tilde{\Omega}_{iv}(\vec{k},\omega)=
	\frac{\Delta ^2}{\Delta ^2+\omega ^2}+\frac{\Delta^2 \left(2 \Delta _2^2+\omega ^2\right)+\Delta_2^2 \omega ^2+12 \Delta  \Delta _2 k^2+8 k^2
	\left(2 k^2+\omega ^2\right)}{(\omega^2+a^{2})(\omega^2+b^{2})}-\frac{8 k^2}{2k^2+\omega ^2}.	
\end{align}
Expanding the middle fraction and performing integration over $\omega$, we find the free energy density for the intervalley gap state
\begin{align}\label{eq:gamma-intervalley}
	&\Omega_{iv}=-\frac{2}{v_F^2}\int\limits_{0}^{\Lambda}\frac{k dk}{2\pi}\left[\frac{|\Delta| }{2}-2\sqrt{2}k+\left(\frac{a^2 \Delta _2^2+\Delta^2 \left(a^2-2
	\Delta _2^2\right)+8 k^2 \left(a^2-2k^2\right)-12 \Delta  \Delta _2k^2}{2a\left(a^2-b^2\right)}+(a\leftrightarrow b)\right)\right].
\end{align}
The free energy density $\Omega$ for the intervalley gap from Sec.\ref{intervalley-numerical} is shown in Fig.\ref{fig:free-energy} by red dashed line.

\end{document}